\documentclass[english]{article}
\usepackage[T1]{fontenc}
\usepackage[latin9]{inputenc}
\usepackage{esint}



\usepackage{babel}
\begin{document}

\title{Algebraic analysis of linear systems with double type of perturbations }

\author{Jerzy Hanckowiak\\Zielona Gora, Poland, EU\\ e-mail: hanckowiak@wp.pl }

\maketitle

\date{September 2013}
\begin{abstract}
Linear systems under the influence of nonlinear and random linear
perturbations, and with random initial and boundary conditions, are
discussed. The notion of states of a system is substituted by the
notion of the generating vectors for n-point information (n-pi), n=0,1,2,...,
characterizing a system at different space-time points. A complete
system of equations for physical n-pi are derived and certain universal
expansions are proposed.

\tableofcontents{}
\end{abstract}

\section{Introduction}

When the external and/or internal, multiplicative and/or additive
random perturbations of a dynamical system appear, denoted symbolically
by $\zeta$, then it is assumed that \textbf{the initial conditions
(IC) cease to have physical value,} see \cite{Kamp 1981}. In this
situation, we are looking for such approximations to solutions of
considered equations in which such dependence on the IC would be suppressed,
or, at least, this would be the case, for appropriately large times,
see again \cite{Kamp 1981} and \cite{Lin 1995,Sob 1991}. 

In this study we realize the above premise by the demand that IC are
also random variables. IC, however, differ from the $\zeta$ that
do not appear in considered equations explicitly. This fact significantly
facilitates the derivation of the equations for the so-called \textbf{n-point
information (n-pi)}, see \cite{Han 2010-1,Monin 1967}. In these equations,
perturbations $\zeta$ of a system are treated as parameters according
to which some sort of averaging or smoothing will be done when equations
for n-pi will be appropriately transformed. For example, equations
for n-pi will be successfully closed (the closure problem). Not to
be underestimated is the fact that the equations for n-pi are linear.
\textbf{This means that after solving the closure problem and after
going to the stochastical equations ($\zeta$ are treated as random
variables) - all the time - we are dealing with linear equations.}
Such is the advantage of using equations in which the IC are treated
as random variables. But that's not all: the linear equation for n-pi
can be formulated in the free Fock space, and this makes the individual
operators appearing in the derived equations to have interesting property:
they are explicitly right-or left-reversible. This opens up access
to the Algebraic Analysis, which provides a number of useful statements
and formulas, see \cite{Prze 1988} and \cite{Han  2013}. To clarify
confusions caused by double meaning of that term, see \cite{Prze 2000}.
It is conceivable that the ease of building a number of inverse operators
is related to the fact that we use generating vectors for multitime
n-pi. This may mean that the use of such vectors is a more appropriate
tool than the use of the concept of the state vectors associated with
a given time. 

It is remarkable that, for the averaged quantities which are the n-pi,
the linear equations are obtained in the case of linear as well as
nonlinear original Eq.7. \textbf{In other words, less detailed description
of the system is expressed by means of linear equations.} A similar
situation can be found in quantum mechanics, see Schrödinger equations.
In the paper, we propose a modification of derived equations for n-pi
by adding a 'quantum term', see Sec.4. It turns out that such a term
can also be obtained in the case of essentially non-linear original
theory (\ref{eq:7}), if in Eq.\ref{eq:8} an appropriate definition
of the operator-valued functions is used, see \cite{Han  2013}. In
a quantum description of systems this operator is responsible for
existing nontrivial perturbative solutions even in the case of non
singular operators $\hat{L}$.

\section{All possible states of a considered system}

Let us assume that all possible states of the system are described
by some function-functional (f-f) $\varphi$ depending on the three
kinds of quantities: $\tilde{x},\zeta,\alpha$:

\begin{equation}
\varphi=\varphi[\tilde{x};\zeta,\alpha]\label{eq:1}
\end{equation}
In Eq.\ref{eq:1}, the components of $\tilde{x}$ describe different
points in the space-time and different components of the field $\varphi$,
$\zeta$- represents different perturbations acting on the system
and $\alpha$ - expresses IC, or/and boundary conditions imposed on
the system. With this interpretation of $\alpha$ we can think that
Eq.\ref{eq:1} expresses the causality relation especially in the
situation in which the $\varphi$ and $\alpha$ belong to the same
type of entities. We used here square brackets to express possible
functional dependence of $\varphi$ on $\zeta$ and $\alpha$. 

What can you say about the system without precise specification of
variables $\tilde{x},\zeta,\alpha$? We can not say that it is a discrete
or continuous system, but we can say that the system is subjected
to some 'unwanted' and possibly random perturbations (forces), which
are denoted by $\zeta$ . If that being the case, the IC (initial
conditions) (and probably the boundary conditions) no longer have
physical meaning, and the same can be said about the various states
of the system. As a particular case of (\ref{eq:1}) is $\varphi$
which does not depend on $\zeta$:

\begin{equation}
\varphi=\varphi[\tilde{x};\alpha]\label{eq:2}
\end{equation}
 But even in this case, the IC, $\alpha$, can also be non-physical
entities when the states described by f-f $\varphi$ are very sensitive
to small changes of $\alpha$. The same can be said when $\zeta$
is not random, but the solutions are very sensitive to small changes
of initial conditions (IC).

In both cases IC,$\alpha$, are treated as a random variable(s) and
one can derive similar equations upon a more physical entities as
the averages, or correlation functions, or other quantities which
we call n-pi. So, in general case, we will consider n-pi:

\begin{equation}
V[\tilde{x}_{(n)};\zeta]\equiv\int\varphi[\tilde{x}_{1};\zeta,\alpha]\cdots\varphi[\tilde{x}_{n};\zeta,\alpha]W[\zeta,\alpha]]\delta[\alpha]\label{eq:3}
\end{equation}
for $n=1,2,...,\infty$, where $W$ is a weighting functional, or
density of probability that the random variable $A$ takes a value$\alpha$
and $\Delta$ takes a value $\zeta$. (They can be functions and so
$\int$ can means the functional integral). $\delta[\alpha]$ is a
generalization of differential $dx$ to the case of function $\alpha$,
see \cite{Rzew 1969}. In fact, one can use a more general integrals
like this:

\begin{equation}
V[\tilde{x}_{(n)};\zeta]=<\varphi[\tilde{x}_{1};\zeta,\cdot]\cdots\varphi[\tilde{x}_{n};\zeta,\cdot]>\equiv\int\varphi[\tilde{x}_{1};\zeta,\alpha]\cdots\varphi[\tilde{x}_{n};\zeta,\alpha]d\mu_{\zeta}[\alpha]\label{eq:4}
\end{equation}
 in which the appropriately defined measure, or pseudo measure, $\mu$,
depends on $\zeta$. 

Now we say something about the equations satisfied by $\varphi$:
The basic idea that we take, and that comes from Newton's is that
their form does not depend on $\alpha$. This which the random variable
$\zeta$ differs from $\alpha$ is that $\zeta$ explicitly enters
the equations, which satisfies $\varphi$. This facts make possible
to derive a complete system of equations for $V[\tilde{x}_{(n)};\zeta]$,
and then we can use the formula

\begin{equation}
V(\tilde{x}_{(n)})=\int V[\tilde{x}_{(n)};\zeta]d\mu[\zeta]\label{eq:5}
\end{equation}
to obtain quantities reflecting in some way our observations. In fact,
sometimes it is assumed that $\zeta$ is a generalized function. In
this case, we will assume that $\zeta$ are the usual functions that
depend on parameter(s) of which we go to the limes after these integrations.
In fact, remarks of App.4 show that such a trick is not necessary.

So, we will assume that IC, $\alpha$, not appear in explicit way
in the considered equation:

\begin{equation}
L[\tilde{x},\zeta;\varphi]+\lambda N[\tilde{x},\zeta;\varphi]+G[\tilde{x},\zeta]=0\label{eq:6}
\end{equation}
Here, $L$ depends linearly on $\varphi$, $N$ - nonlinear, and $G$
does not depend on $\varphi$ at all. $\lambda$ is an expansion parameter
describing the strength of nonlinear perturbation of the linear (kinematic)
theory. 

We assume that dependence on the random perturbations $\zeta$ in
Eq.\ref{eq:6} is linear. Therefore, we consider equation:

\begin{equation}
L[\tilde{x};\varphi]+\lambda N[\tilde{x};\varphi]+G(\tilde{x})+\zeta(\tilde{x}')\Delta[\tilde{x}';\varphi]=0\label{eq:7}
\end{equation}
In the last term we understand summation with respect to certain components
of the vector $\tilde{x}'$. We remind you that in the previous author's
papers, to save space, we put in $\tilde{x}$ the discrete indexes
of the field $\varphi$ . We can do the same, for the perturbation
field $\zeta$. The $\tilde{x}'$ means that there are additional
indexes in comparison with $\tilde{x}$ and the summation in the last
part of the Eq.\ref{eq:7} applies to these indexes. 

For $\Delta=const$ in $\varphi$, we have \textit{additive perturbations}.
In other case, we have \textit{multiplicative perturbations. }In fact,
$\Delta$ can be proportional to $L,N,G$ and to another quantity.
In the paper, we will assume, for $\Delta$, a linear dependence on
$\varphi$ and certain quantum mechanical generalization. 

Eq.$7$ describes micro-properties of the system, which are too detailed.
Less detailed information about the system are described by the linear
equation 

\begin{equation}
\left(\hat{L}+\lambda\hat{N}+\hat{G}+\zeta\cdot\hat{\Delta}\right)|V;\zeta>=|0;\zeta>_{info}=\hat{P}_{0}|0;\zeta>_{info}\label{eq:8}
\end{equation}
for the vector $|V;\zeta>$ generating n-pi given by the formulas
(\ref{eq:3}) or (\ref{eq:4}), see also Eq.\ref{eq:9} and Eq.\ref{eq:13}.
A very simple derivation of this equation and explicit forms of operators
$\hat{L},\hat{N},\hat{G}$, and definition of 'vacuum' vector $|0;\zeta>_{info}$,
can be found in \cite{Han 2011}. See, however, App.1 where it is
shown that in a \textbf{certain cases an appropriate set of 'less
detailed information' about the system may be used to retrieve detailed
information}. 

Let us notice the  difference between meaning of symbols $N$ and
$\hat{N}$!! The first means nonlinear functional, the second means
a linear operator which is closely related to the first. 

Uncontrolled perturbations $\zeta$ appearing in this equation can
be treated as a fixed quantity under which we will make averaging
when we find a solution to Eq.$\ref{eq:8}$ represented by the series:

\begin{equation}
|V;\zeta>=\sum_{n=1}\int d\tilde{x}_{(n)}V[\tilde{x}_{(n)};\zeta]|\tilde{x}_{(n)}>+|0;\zeta>_{info}\label{eq:9}
\end{equation}
with the n-point functions $V[\tilde{x}_{(n)};\zeta]$ called the
\textit{n-point information} \textit{(n-pi) about the system}, which
have interpretation given by Eqs (\ref{eq:3}) or (\ref{eq:4}). The
vectors $|\tilde{x}_{(n)}>$, for n=1,2..., are linearly independent
orthonormal vectors explicitly expressed by Eqs (\ref{eq:13}). The
vector $|0;\zeta>_{info}$ describes so called the \textit{local information
vacuum, }see \cite{Han 2010} and \cite{Han 2012}, and as the only
basis vectors does not depend on the $\tilde{x}$, see also App.2. 

The operator $\hat{L}$ is a right invertible operator, which in the
case of \textit{classical statistical field theory} is a diagonal
operator with respect to the projectors $\hat{P}_{n};n=1,2,...,\infty$:

\begin{equation}
\hat{P}_{n}\hat{L}=\hat{L}\hat{P}_{n}\label{eq:10}
\end{equation}
 where the project $\hat{P}_{n}$projects on the n-th term in the
expansion (\ref{eq:9}). The projector $\hat{P}_{0}$projects on the
subspace generated by the vector $|0;\zeta>_{info}$. 

In the case of \textit{quantum field theory,} the operator $\hat{L}$
is an invertible or right invertible diagonal, plus a lower triangular
operator related to the commutation relations of the canonical conjugate
operator variables, with respect to the same set of projectors $\hat{P}_{n}$. 

In the case of \textbf{polynomial} nonlinearity, the operator $\hat{N}$
is an upper triangular operator in a classical as well as in quantum
field theory:

\begin{equation}
\hat{P}_{n}\hat{N}=\sum_{n<m}\hat{P}_{n}\hat{N}\hat{P}_{m}\label{eq:11}
\end{equation}
 see \cite{Han 2012}. 

The operator $\hat{G}$ , in the both cases, is a left invertible
operator, which is lower triangular operator:

\begin{equation}
\hat{P}_{n}\hat{G}=\sum_{m<n}\hat{P}_{n}\hat{G}\hat{P}_{m}\label{eq:12}
\end{equation}
 Similar property has the operator $\hat{\Delta}$ discussed in Sec.4. 

All these operators are linear operators considered in the \textbf{\textit{free
Fock space }}(FFS) constructed by means of the vectors like (\ref{eq:9})
in which

\begin{equation}
|\tilde{x}_{(n)}>=\hat{\eta}^{\star}(\tilde{x}_{1})\cdots\hat{\eta}^{\star}(\tilde{x}_{n})|0>\label{eq:13}
\end{equation}
and the operators $\hat{\eta}^{\star}$ satisfy the Cuntz relations:

\begin{equation}
\hat{\eta}(\tilde{y})\hat{\eta}^{\star}(\tilde{x})=\delta(\tilde{y}-\tilde{x})\cdot\hat{I}\label{eq:14}
\end{equation}
 where $\left(\hat{\eta}^{\star}(\tilde{x})\right)^{\star}=\hat{\eta}(\tilde{x})$
, $\hat{I}$ is the unit operator in $FFS$ and other relations take
place:

\begin{equation}
\hat{\eta}(\tilde{y})|0>=0,\quad<0|\hat{\eta}^{\star}(\tilde{y})=0\label{eq:15}
\end{equation}
see \cite{Han 2012}. Moreover, we will assume that all components
of vectors $\tilde{x},\tilde{y}$ are discrete - so that $\delta(\tilde{y}-\tilde{x})$,
in fact, is Kronecker delta. The operator $\zeta\cdot\hat{\Delta}$
will be described in Secs 4 and 5.

\section{A dominant role of kinematic term $\hat{L}$}

In this case a solution is sought in the form of series:

\begin{equation}
|V;\zeta>=\sum_{j=0}^{\infty}\lambda^{j}|V;\zeta>^{(j)}\label{eq:16}
\end{equation}
To get subsequent approximations, $\lambda^{j}|V;\zeta>^{(j)}$, we
transform Eq.\ref{eq:8} as follows: we will assume that the kinematic
term $\hat{L}$is a right invertible operator:

\begin{equation}
\hat{L}\hat{L}_{R}^{-1}=\hat{I}\label{eq:17}
\end{equation}
 We get then the equivalent equation to the Eq.\ref{eq:8}:

\begin{eqnarray}
 & \left(\hat{I}+\hat{L}_{R}^{-1}\left(\lambda\hat{N}+\zeta\cdot\hat{\Delta}\right)+\hat{L}_{R}^{-1}\hat{G}\right)|V;\zeta & >=\nonumber \\
 & \hat{L}_{R}^{-1}|0>_{info}+\hat{P}_{L}|V;\zeta>\label{eq:18}
\end{eqnarray}
 where a projector

\begin{equation}
\hat{P}_{L}=\hat{I}-\hat{L}_{R}^{-1}\hat{L}\label{eq:19}
\end{equation}
 Taking into account the permutation symmetry condition:

\begin{equation}
|V;\zeta>=\hat{S}|V;\zeta>,\label{eq:20}
\end{equation}
 (for an explicit form of the projector $\hat{S}$ see \cite{Han 2007}),
we can project Eq.\ref{eq:18} as follows:

\begin{eqnarray}
 & \left(\hat{I}+\hat{S}\hat{L}_{R}^{-1}\left(\lambda\hat{N}+\zeta\cdot\hat{\Delta}\right)+\hat{S}\hat{L}_{R}^{-1}\hat{G}\right)|V;\zeta>=\nonumber \\
 & \hat{S}\hat{L}_{R}^{-1}|0;\zeta>_{info}+\hat{S}\hat{P}_{L}|V;\zeta>\label{eq:21}
\end{eqnarray}
 This equation differs from Eq.\ref{eq:18} that an arbitrary part
of the solution $|V;\zeta>$, $\hat{S}\hat{P}_{L}|V;\zeta>$, is projected
on the smaller part of FFS than in the case of Eq.\ref{eq:18}. We
will identify this part with the zero-th approximation $(\lambda=0)$:

\begin{equation}
\hat{S}\hat{P}_{L}|V;\zeta>=\hat{S}\hat{P}_{L}|V;\zeta>^{(0)}\label{eq:22}
\end{equation}
 To get higher approximation terms in the expansion (\ref{eq:16}),
it is recommended to transform Eq.\ref{eq:21} as follows:

\begin{eqnarray}
 & \left\{ \hat{I}+\lambda\left[\hat{I}+\hat{S}\hat{L}_{R}^{-1}\left(\hat{G}+\zeta\cdot\hat{\Delta}\right)\right]^{-1}\hat{S}\hat{L}_{R}^{-1}\hat{N}\right\} |V;\zeta>=\nonumber \\
 & \left[\hat{I}+\hat{S}\hat{L}_{R}^{-1}\left(\hat{G}+\zeta\cdot\hat{\Delta}\right)\right]^{-1}\left(\hat{S}\hat{L}_{R}^{-1}|0;\zeta>_{info}+\hat{S}\hat{P}_{L}|V;\zeta>\right)\label{eq:23}
\end{eqnarray}
 Are the equations (\ref{eq:21}) and (\ref{eq:23}) equivalent to
the previous equations? The answer may be positive, if the previous
equations are overdetermined equations. There are two reasons for
this: considered previous equations have identical shape in the free
(FFS) as well as in the symmetrical Fock space (SFS), and another
more fundamental and surprising reason is such that Eq.\ref{eq:8}
can be overdetermined even with respect to the micro-equation (\ref{eq:7}),
see App.1. 

For lower triangular perturbation operators $\hat{\Delta}$, see Sec.4,
the inverse appearing in Eq.\ref{eq:23} is not difficult to calculate,
for any values of the $\zeta$. There is another problem related to
these operators, namely - we can not be sure that the positivity conditions
considered in Sec.5 are satisfied in the case when the term (\ref{eq:25})
occurs in the perturbation operator$\hat{\Delta}.$ However, the functional
integration representation of solutions convinces us that it is not
the case. 

The expansion (\ref{eq:16}) is useful and in some sense obligatory,
for an infinite system of branching equations. However, for essential
nonlinearities considered in \cite{Han  2013}, obtained equations
for n-pi, are closed. In this case, one can find a more effective
(e.g. numeriacal) methods of approximations to the generating vectors
$|V;\zeta>$. Moreover, equations for $|V>$can be easy derived, see
App.4.

\section{Random perturbations $\zeta\cdot\hat{\Delta}$}

Until now we did not say anything about the \textit{random perturbation
operator,} $\zeta\cdot\hat{\Delta}$. We know that in FFS n-pi are
described by Eq.\ref{eq:8} in which external forces acting on the
system are described by the lower triangular operator

\begin{equation}
\hat{G}_{ext}=\sum_{n=1}^{\infty}\hat{P}_{n}\hat{G}_{ext}\hat{P}_{n-1}\label{eq:24}
\end{equation}
see \cite{Han 2011'}. On the other hand, we know that the symmetric
Green's functions of quantum field theory, which codify the causality
condition, are described by equations of the type (\ref{eq:8}), where
$G_{QFT}$ is the operator with property:

\begin{equation}
\hat{G}_{QFT}=\sum_{n=2}^{\infty}\hat{P}_{n}\hat{G}_{QFT}\hat{P}_{n-2}\label{eq:25}
\end{equation}
 The latter operator describes disorders of the system caused by attempting
simultaneous measurement of canonically conjugate variables. So, we
propose the following linear combination

\begin{equation}
\zeta\cdot\hat{\Delta}=\zeta_{exe}\cdot\hat{G}_{ext}+\zeta_{QFT}\cdot\hat{G}_{QFT}\label{eq:26}
\end{equation}
for the operator $\zeta\cdot\Delta$. In this way we can hope that
the above formula gives a general frame to describe a system under
influence (perturbations) of the random external fields created by
the entities simultaneously subjected to quantum fluctuations. In
the case of the system of Brownian particles (dust particles), the
random perturbations (fluctuations) are caused by fast moving atoms
in the gas or liquid. In the case of the economic system, the 'dust
particles' can be large companies and atoms may be substituted by
smaller units like customers.

\section{Positivity conditions and the term $\zeta_{QFT}\cdot\hat{G}_{QFT}$}

Eq.\ref{eq:8} derived by means of definitions of n-pi (\ref{eq:4})
and Eq.\ref{eq:7} allows operators which are diagonal, upper triangular
and very specific lower triangular operators as (\ref{eq:24}). But
it seems the operator $\zeta_{QFT}\cdot\hat{G}_{QFT}$ does not belong
to these classes of operators. This can causes a conflict with the
positivity conditions for 2-pi $V[\tilde{x},\tilde{y};\zeta]$ resulting
from the definitions ($\ref{eq:3})$ or (\ref{eq:4}):

\begin{equation}
\sum_{i,j=1}^{n}V[\tilde{x_{i}},\tilde{x}_{j};\zeta]\eta(\tilde{x}_{i})\eta(\tilde{x}_{j})\geq0\label{eq:27}
\end{equation}
 for an arbitrary choice of $n$, functions $\eta,\zeta$ and points
$\tilde{x}_{i},\tilde{y}_{j}$. (We assume that considered fields
$\varphi,\eta,\zeta$ are real-valued functions). Please do not confuse
$\eta$ without the hat with $\hat{\eta}$ with the hat, which is
the operator satisfying the Cuntz relations. Similar restrictions
also occur for higher even n-pi. See also \cite{Han 2010}, Sec.3.
After integration of inequality (\ref{eq:27}), the same inequality
is satisfied by the functions (\ref{eq:5}). 

One can prove at least formally, by means of the functional integral
representation, see e.g. \cite{Rzew 1969}, that positivity conditions
are also satisfied in the case of Eq.\ref{eq:8} with terms given
by Eq.\ref{eq:25}. To see this let us define the n-pi with the help
of generating functional integral:

\begin{equation}
V[\eta;\zeta]=\int\delta\beta\, e^{-H[\beta;\zeta]}e^{i\beta\eta}\label{eq:29-1}
\end{equation}
 where $\eta,\beta,\zeta$ are real functions and $\beta\eta$ in
the second exponent - a scalar product, see \cite{Rzew 1969}. It
is easy to see that n-pi defined by the n-th order functional derivatives:

\begin{equation}
-(i)^{n}\frac{\delta^{n}}{\delta\eta(\tilde{y}_{1})\cdots\delta\eta(\tilde{y}_{n})}V[\eta;\zeta]_{\eta=0}\equiv V[\tilde{x}_{(n)};\zeta]\label{eq:30-1}
\end{equation}
 satisfy the above positivity conditions if $H$ is a real function.
One can also show that the generating functional (\ref{eq:29-1})
satisfies an equation similar to Eq.\ref{eq:8}, see \cite{Rzew 1969}. 

In the case of \textit{highly nonlinear interaction} $N$, in Eq.\ref{eq:7},
(by this we mean a non-polynomial dependence of the $N$ functional
on the field $\varphi$ one can define the linear operator $\hat{N}$
of Eq.\ref{eq:8}, which in addition to the diagonal part contains
the lower triangular part, see \cite{Han  2013}; Sec.2. \textbf{One
can use positivity conditions (\ref{eq:27}), and similar relations,
to justify for certain values of parameters}, at least,\textbf{ proposed
definitions of operator-valued functions}.

\section{App.1 What can you say about solutions to an equation with the knowledge
of their averages and n-point information (n-pi)?}

We want to give a simple example that the knowledge of averages of
'fields' $\varphi[\tilde{x};\alpha]$ and their n-pf (e.g., correlation
functions) allows to reconstruct the fields. We consider extremely
simple case in which $\tilde{x},\alpha$ take only two values: 1 and
2. We take arithmetic averages, which correspond to a constant probability
equal in this case to 1/2. So, we have:

\begin{equation}
<\varphi[\tilde{x};\cdot]>=1/2\left(\varphi[\tilde{x};1]+\varphi[\tilde{x};2]\right)\label{eq:28}
\end{equation}
 for , (Two equations). For correlation functions (2-pi) , we get
additional 4 equations: 

\begin{equation}
<\varphi[\tilde{x}_{i};\cdot]\varphi[\tilde{y}_{j};\cdot]>=1/2\left(\varphi[\tilde{x}_{i};1]\varphi[\tilde{y}_{j};1]+\varphi[\tilde{x}_{i};2]\varphi[\tilde{y}_{j};2]\right)\label{eq:29}
\end{equation}
 by means of which one can calculate the field $\varphi[\tilde{x};\alpha]$
. We assume here that the l.h.s. of the above equations are known;
The equation which satisfies the field together with the additional
conditions allow to find them in a unique way. 

In other words, it is possible that there are situations in which
averaged formulas are used because they - in contrary to unaveraged
quantities - have got the physical meanings.

\section{App.2 About vectors $|0;\zeta>_{info}$and |0>}

Why in the Eq.\ref{eq:8} appears the vectors $|0;\zeta>_{info}$?
This vector appears in Eq.\ref{eq:8} because we want to have equations
for the generating vector $|V>$with right inverse operators defined
in the whole FFS, see \cite{Han 2010}. Of course, appropriate modifications
of Eq.\ref{eq:8}, for generating vector $|V>$, is made in accordance
with the definitions of n-pi and micro- (local) Eq.\ref{eq:7}. It
results, from construction of vector $|0>_{info}$ in \cite{Han 2011'},
that 

\begin{equation}
|0;\zeta>_{info}\sim|0>\label{eq:30}
\end{equation}
In \cite{Han 2012} we assumed that $|0;\zeta>_{info}$ represents
the local vacuum (no local information about the system). It depends
on the global characteristics of the system and perturbations $\zeta$:

\begin{equation}
|0;\zeta>_{info}=|0;V,\zeta>_{info}\label{eq:31}
\end{equation}
 The system and random perturbations acting on the system, both form
a quasi-isolated system, for which the rest of the world, with a good
approximation, can be treated as the whole Universe

In \cite{Han 2012} we have identified the basic vector $|0>$ of
FFS with a vector $|U>$ describing the whole Universe:

\begin{equation}
|0>=|U>\label{eq:32}
\end{equation}
 The reason may be this that with the help of this vector, for the
most complex systems, a whole Fock space can be created. For example
- the Universe. With the help of this vector and creation operators
$\hat{\eta}^{\star}(\widetilde{x})$ one can retrieve all local information
about the system

\begin{eqnarray}
 & |V>=\int|V;\zeta>d\mu[\zeta]=\nonumber \\
 & \sum_{n=1}\int d\tilde{x}_{(n)}\left\{ \left(\int V[\tilde{x}_{(n)};\zeta]d\mu[\zeta]\right)\hat{\eta}^{\star}(\tilde{x}_{1})\cdots\hat{\eta}^{\star}(\tilde{x}_{n})|U>\right\} +|0;V>_{info}\label{eq:34-1}
\end{eqnarray}
where 

\begin{equation}
|0;V>_{info}=\int d\mu[\zeta]|0;\zeta>_{info}\equiv|0>_{info}\label{eq:34}
\end{equation}

\section{App.3 About creation and annihilation operators $\hat{\eta}^{\star}(\tilde{x})$
and $\hat{\eta}(\tilde{x})$}

These operators can depend on considered systems, but in calculating
the n-pi we use only Cuntz relations (\ref{eq:14}) and (\ref{eq:15}).
You do not even need to use an involutional property: xxx

\begin{equation}
\left(\hat{\eta}^{\star}\right)^{\star}=\hat{\eta}\label{eq:35}
\end{equation}

\section{App.4 About equations for n-pi $V(\tilde{x}_{(n)})$ generated by
the vector |V>}

By straight integration, it is seen from Eq.\ref{eq:23} that the
j-th order approximation to n-pi $V(\tilde{x}_{(n)})$ is given by
the lower order approximations to m-pi $V[\tilde{x}_{(m)};\zeta]$
generated by the vector $|V;\zeta>$. 

For closed equations for n-pi, obtained in the case of essentially
nonlinear theories, the n-pi $V(\tilde{x}_{(n)})$ is expressed by
m-pi $V[\tilde{x}_{(m)};\zeta]$, with $m<n$. In other words, even
when we do not have exclusive (complete) equations for physical n-pi
$V(\tilde{x}_{(n)})$, there calculations are realized by \textit{simpler},
in the above sense, m-pi $V[\tilde{x}_{(m)};\zeta]$. 

In general case, by integrating Eq.\ref{eq:8} with respect to variable
$\zeta$, we get the following equation for generating vector $|V>$:

\begin{eqnarray}
 & \left[\left(\hat{L}+\lambda\hat{N}+\hat{G}\right)|V>+\int d\mu[\zeta]\star\zeta\cdot\hat{\Delta}|V;\zeta>\right]=\nonumber \\
 & \int d\mu[\zeta]|0;\zeta>_{info}=\hat{P}_{0}\int d\mu[\zeta]|0;\zeta>_{info}\label{eq:37}
\end{eqnarray}
 see (\ref{eq:3})-(\ref{eq:5}). Of course, in this equation we have
to know the generating vector $|V;\zeta>$ and the local vacuum vector
$|0;\zeta>_{info}$, for a set of all possible perturbations $\zeta$.
However, we do not need to know these quantities in a very precise
manner because they are smoothed out by integration. We can go back
to Eq.\ref{eq:8} and seek solutions, for $|V;\zeta>$, in the form
of a Volterra series, e.g.,:

\begin{equation}
|V;\zeta>=\sum_{j=0}1/j!\int d\tilde{y}_{(j)}|V(\tilde{y}_{(j)})>\zeta(\tilde{y}_{1})\cdots\zeta(\tilde{y}_{j})\label{eq:38}
\end{equation}
 see \cite{Rzew 1969}. Here

\begin{equation}
|V(\tilde{y}_{(j)})>_{pertur}=\frac{\delta^{j}}{\delta\zeta(\tilde{y}_{1})\cdots\delta\zeta(\tilde{y}_{j})}|V;\zeta>|_{\zeta=0}\label{eq:39}
\end{equation}
 Equations for vectors $|V(\tilde{y}_{(j)})>$ are complete, see Eq.\ref{eq:8}.
We must note here that if the perturbation field $\zeta$ contains
any sub-indices, then the same sub-indices are included in vectors
$|V(\tilde{y}_{(j)})>$ and then in the formula (\ref{eq:38}) we
use Einstein's summation convention. 

We have the following relations:

\begin{eqnarray}
 & |V;\zeta>=\sum_{j=0}^{\infty}1/j!\int d\tilde{y}_{(j)}|V(\tilde{y}_{(j)})>_{pertur}\zeta(\tilde{y}_{1})\cdots\zeta(\tilde{y}_{j})=\nonumber \\
 & \sum_{n=1}\int d\tilde{x}_{(n)}V[\tilde{x}_{(n)};\zeta]\hat{\eta}^{\star}(\tilde{x}_{1})\cdots\hat{\eta}^{\star}(\tilde{x}_{n})|0>+|0;\zeta>_{info}\label{eq:40}
\end{eqnarray}
 From the Cuntz relations (\ref{eq:14})-(\ref{eq:15}), we get:

\begin{eqnarray}
 & V[\tilde{x}_{(n)};\zeta]=<\hat{\eta}(\tilde{x}_{1})\cdots\hat{\eta}(\tilde{x}_{n})|V;\zeta>=\nonumber \\
 & \sum_{j=0}^{\infty}\frac{1}{j!}\int d\tilde{y}_{(j)}<\hat{\eta}(\tilde{x}_{1})\cdots\hat{\eta}(\tilde{x}_{n})|V(\tilde{y}_{(j)})>_{pertur}\zeta(\tilde{y}_{1})\cdots\zeta(\tilde{y}_{j})\label{eq:41}
\end{eqnarray}
and from (\ref{eq:5}):

\begin{eqnarray}
 & V(\tilde{x}_{(n)})=\int V[\tilde{x}_{(n)};\zeta]d\mu[\zeta]=\nonumber \\
 & \sum_{j=0}^{\infty}\frac{1}{j!}\int d\tilde{y}_{(j)}<\hat{\eta}(\tilde{x}_{1})\cdots\hat{\eta}(\tilde{x}_{n})|V(\tilde{y}_{(j)})>_{pertur}\int d\mu[\zeta]\zeta(\tilde{y}_{1})\cdots\zeta(\tilde{y}_{j})\label{eq:42}
\end{eqnarray}
 where, in this and other equations, we freely have changed the orders
of appropriate operations.

In the case of small perturbations $\zeta$, only few terms of the
Volterra series (\ref{eq:42}) has to be taken into account:-) to
get physical n-pi $V(\tilde{x}_{(n)})$. Hence, we claim that equations
for $V(\tilde{x}_{(n)})$ are also complete. We do not take into account
the $\zeta$- dependence of the vacuum vector $|0;\zeta>_{info}$
which indeed does not appear, for certain, perturbative type calculations.

\section{App.5 Difference between completeness and closeness of equations}

Equations, for a given \textbf{set} of quantities are called complete
if by means of them, with the help of reasonable set of additional
conditions, one can solve them.

Equations, for a given \textbf{subset} of quantities are called close
if by means of them, with the help of reasonable set of additional
conditions, one can solve them. In this way the \textsl{closure problem}
is solved.


\begin{thebibliography}{10}
\bibitem[1]{Kamp 1981}van Kampen, N.G. 1981. \textit{Stochastic processes
in physics and chemistry. }Elsevier Science Publishers B.V. 1981

\bibitem[2]{Lin 1995}Lin, Y.K. and G.Q. Cai. 1995. \textit{Probabilistic
Structural Dynamics. }McGraw-Hill, New York.

\bibitem[3]{Sob 1991}Sobczyk, K. 1991. \textit{Stochastic Differential
Equations with Applications to Physics and Engimeering. }Kluwer Academic
Publishers B.V. 1991

\bibitem[4]{Monin 1967}Monin, A.S. and A.M. Jaglom. 1967. \textit{Statistical
Hydromechanics, vol. 2, }MIR Publishers, Moscow.

\bibitem[5]{Han 2010-1}Hanckowiak, J. 2010. \textit{Models of the
'Universe' and a Closure Principle. }arXiv: 1010.3352 physics.gen-ph

\bibitem[6]{Han  2013}Hanckowiak, J. 2013. \textit{Nonlinearity and
linearity, friends or enemies? Algebraic Analyzation of Science:).
}arXiv:1304.3453v1 {[}physics.gen-ph{]} 11 Apr 2013

\bibitem[7]{Han 2011}Hanckowiak, J. 2011. \textit{Unification of
some classical and quantum ideas, }arXiv:1107.1365v1 

\bibitem[8]{Han 2012}Hanckowiak, J. 2012. \textit{Metaphysics of
the free Fock space with local and global information. }arXiv:1206.4589v1{[}physics.gen-ph{]} 

\bibitem[9]{Han 2011'}Hanckowiak, J. 2011'. \textit{Local and global
information and equations with left and right invertible operators
in the free Fock space. }arXiv:1112.1870v1 {[}physics.gen-ph{]} 

\bibitem[10]{Prze 1988}Przeworska-Rolewicz, D. 1988. \textit{Algebraic
Analysis. }PWN Polish Science Publishers, Warsaw and D. Reidel Publishing
Company. Dordercht..., Tokyo

\bibitem[11]{Prze 2000}Przeworska-Rolewicz, D. 2000. \textit{Two
centuries of the term 'Algebraic Analysis'. }PAS, Warsaw (Internet)

\bibitem[12]{Han 2010}Hanckowiak, J. 2010. \textit{Free Fock space
and functional calculus approach to the n-point information about
the 'Universe'. }arXiv:1011.3250v1 {[}physics.gen-ph{]}

\bibitem[13]{Rzew 1969}Rzewuski, J. 1969. \textit{Field Theory, part
II. }Ilife Books LTE, London

\bibitem[14]{Han 2007}Hanckowiak, J. 2007. \textit{Reynolds' dream?
A description of random field theory within a framework of algebraic
analysis and classical mechanics. }Nonlinear Dyn. (2007). 50. 191-211\end{thebibliography}
\end{document}